\documentclass[aps,prb,twocolumn,amsmath,amssymb,showpacs,prl,unsortedaddress]{revtex4-1} 
\usepackage{epsf}      
\usepackage{graphicx}
\usepackage{gensymb}
\usepackage{sidecap}
\usepackage{color}
\usepackage{textcomp}

\makeatletter

\usepackage{url}
\usepackage{graphicx}

\@ifundefined{textcolor}{}
{
 \definecolor{BLACK}{gray}{0}
 \definecolor{WHITE}{gray}{1}
 \definecolor{RED}{rgb}{1,0,0}
 \definecolor{GREEN}{rgb}{0,1,0}
 \definecolor{BLUE}{rgb}{0,0,1}
 \definecolor{CYAN}{cmyk}{1,0,0,0}
 \definecolor{MAGENTA}{cmyk}{0,1,0,0}
 \definecolor{YELLOW}{cmyk}{0,0,1,0}
}

\begin{document}

\title {Imaging the magnetic nano-domains in Nd$_2$Fe$_{14}$B}

\author{Lunan Huang}
\author{Valentin Taufour}
\author{T. N. Lamichhane}
\author{Benjamin Schrunk}
\author{Sergei L. Bud'ko}
\author{P. C. Canfield}
\author{Adam Kaminski}
\affiliation{Division of Materials Science and Engineering, The Ames Laboratory, U.S. DOE}
\affiliation{Department of Physics and Astronomy, Iowa State University, Ames, Iowa 50011, USA}

\date{\today}                                         

\begin{abstract}
We study magnetic domains in Nd$_2$Fe$_{14}$B single crystals using high resolution magnetic force microscopy (MFM). Previous MFM studies and small angle neutron scattering experiments suggested the presence of nano-scale domains in addition to optically detected micrometer-scale ones. We find, in addition to the elongated, wavy nano-domains reported by a previous MFM study, that the micrometer size, star shape fractal pattern is constructed of an elongated network of nano-domains $\sim$20 nm in width, with resolution-limited domain walls thinner than 2 nm. While the microscopic domains exhibit significant resilience to an external magnetic field,  some of the nano domains are sensitive to the magnetic field of the MFM tip. 
\end{abstract}

\maketitle

\section{Introduction}

Neodymium iron boron (Nd$_2$Fe$_{14}$B) was developed in 1982 by General Motors and Sumitomo Special Metals and is one of the most popular magnetic materials for advanced applications. It is used in a variety of devices ranging from actuators, high capacity hard drives, to lightweight, high efficiency electric motors for cars. Nd$_2$Fe$_{14}$B is one of the strongest permanent magnets known. During the past three decades, many studies have studied its properties. More recent studies have typically focused on the development of materials with similar magnetic properties that do not require the use of rare earths elements. To accomplish this one needs to fully understand the physical mechanism that gives rise to the unusually enhanced magnetic properties of this material. Nd$_2$Fe$_{14}$B has a tetragonal lattice symmetry with 68 atoms per unit cell. The lattice constants\cite{Herbst91} are a = 8.80 \AA{}, c = 12.20 \AA{}. It has a Curie temperature of 565~K \cite{Buschow86,Herbst91} and a spin-reorientation temperature of $T_{SR}$ = 135~K. Between these temperatures its magnetic moment is aligned along the c axis and below 135~K the alignment depends on temperature. At 4~K the magnetic moment has an angle of about 30\textdegree from the c axis toward the [110] direction \cite{Herbst91}.

The magnetic domain structure of Nd$_2$Fe$_{14}$B has been studied by Lorentz transmission electron microscopy (TEM) and magnetic force microscopy (MFM) in thin film and polycrystalline samples\cite{Lemke97, Grutter88, Neu04, Szmaja06}. Electron microscopy\cite{Lewis1, Lewis2}, Kerr optical microscopy, small angle neutron scattering \cite{Kreyssig09} and MFM \cite{Al-Khafaji98} studies were also carried out using single crystals. These studies reveal that the magnetic structure consists of intriguing fractal patterns that depend on the sample treatment and temperature\cite{Pastushenkov97, Al-Khafaji98, Kreyssig09}. Previous MFM  \cite{Al-Khafaji98} and small angle scattering studies\cite{Kreyssig09} indicated the presence of an even smaller, sub-domain magnetic structure with a typical length scale of 25 - 100 nm. At room temperature the microscopic magnetic domains form a star-like pattern, while below $T_{SR} \sim$ 100~K, they become rectangular in shape. In both temperature regimes, the magnetic domains are arranged in chains\cite{Kreyssig09}. A detailed study of the magnetic domains in this material is interesting from the point of view of fundamental physics as well as practical applications. In this paper we study the morphology of the nano-domains in detail by using high resolution MFM. We find that the star structure present at room temperature is formed from a complicated network of elongated domains with typical widths of $\sim$20nm. The domain walls are even thinner, with a width that is limited by our experimental resolution of 2 nm.

\section{Experimental details}

The Nd${_2}$Fe$_{14}$B crystals were grown out of a Nd-rich ternary melt as in refs.~\onlinecite{Kreyssig09, Lewis2} using a 3-cap, Ta crucible \cite{crucible}. The starting composition of Nd$_{53}$Fe$_{45}$B$_2$ was placed, in elemental form, in the crucible and heated to 1175 C then it was cooled over 105 hours to 800 C.  At this stage, the excess liquid was separated from the plate-like single crystals.

The as-grown, single crystals have flat, shiny facets of nearly optical quality. However, a thin layer of flux binds small particulates with a significant surface density. These particulates interfere with the cantilever and often produce extrinsic magnetic gradients that obscure the MFM signal. To avoid this problem single crystals with a typical size of 5-10 mm were cut to 1 mm thin slices by a low speed diamond wheel. Their surfaces were carefully mechanically polished using powdered alumina with decreasing grain size from 10 to 0.05 $\mu$m yielding a typical surface roughness that is better than 10 nm. After polishing, the sample surface was cleaned with acetone and ethanol and mounted on the sample plate. The measurements were carried out using a Variable Temperature, UHV Scanning Probe Microscope made by Omicron. The surface topography was measured using a non-magnetic AFM cantilever in non-contact mode with a force constant of 42N/m, resonance frequency of 320kHz and reference frequency of 511 kHz. The force between the sample surface and tip was measured via changes in the oscillation frequency of the cantilever and detected optically by a laser deflection sensor. The magnetic structure at the surface was measured using a super sharp silicon, high resolution MFM tip, which has a layer of hard magnetic coating with coercivity of app. 125 Oe and a remanence magnetization of app. 80 emu/cm$^3$ (SSS-QMFMR made by Nano-world). The tip has a force constant of 2.8 N/m and a radius that is less than 15 nm. The force that acts on the magnetized tip depends on the gradient of the magnetic field near the tip. This force changes the oscillating frequency of the cantilever and is detected as described above. The change in frequency of the cantilever oscillation is therefore a measure of the magnetic field gradient at a given point. 

To estimate the roughness of the surface, we imaged the topography of the sample surface using a non-magnetic tip in non-contact mode (shown in Fig.1).  The measurement is performed with the tip traveling very close (a few angstroms) to the sample surface. The roughness of the surface after polishing is approx. 18nm and all features are very irregular. Since the magnetic imaging is performed at a much larger tip to surface distance (100's of nanometers) this level of sample roughness does not significantly affect our measurements.

\section{Results and Discussion} 

\begin{figure}
\includegraphics[width=3.2in]{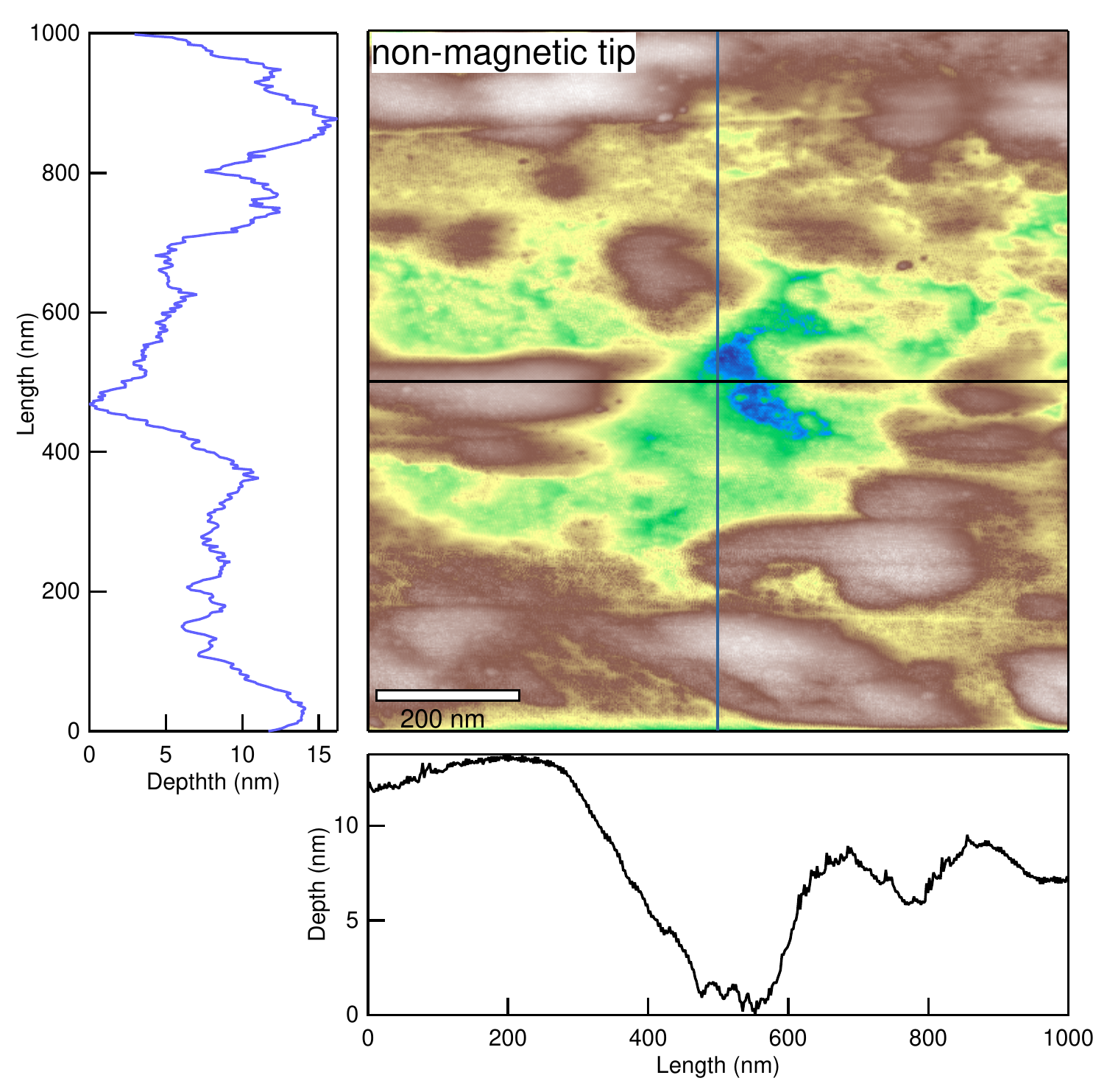}
\caption{(Color online) Main: The surface topographic image of Nd$_2$Fe$_1$$_4$B, non-magnetic signal scanned by the AFM Non-Contact-tip, z mode. 1 $\mu$m * 1 $\mu$m scan, the z range is 18 nm. Left: The z-profile of the vertical line in main. Bottom: The z-profile of the horizontal line in main.}
\label{topographyScan}
\end{figure}

\begin{figure}
\includegraphics[width=3.2in]{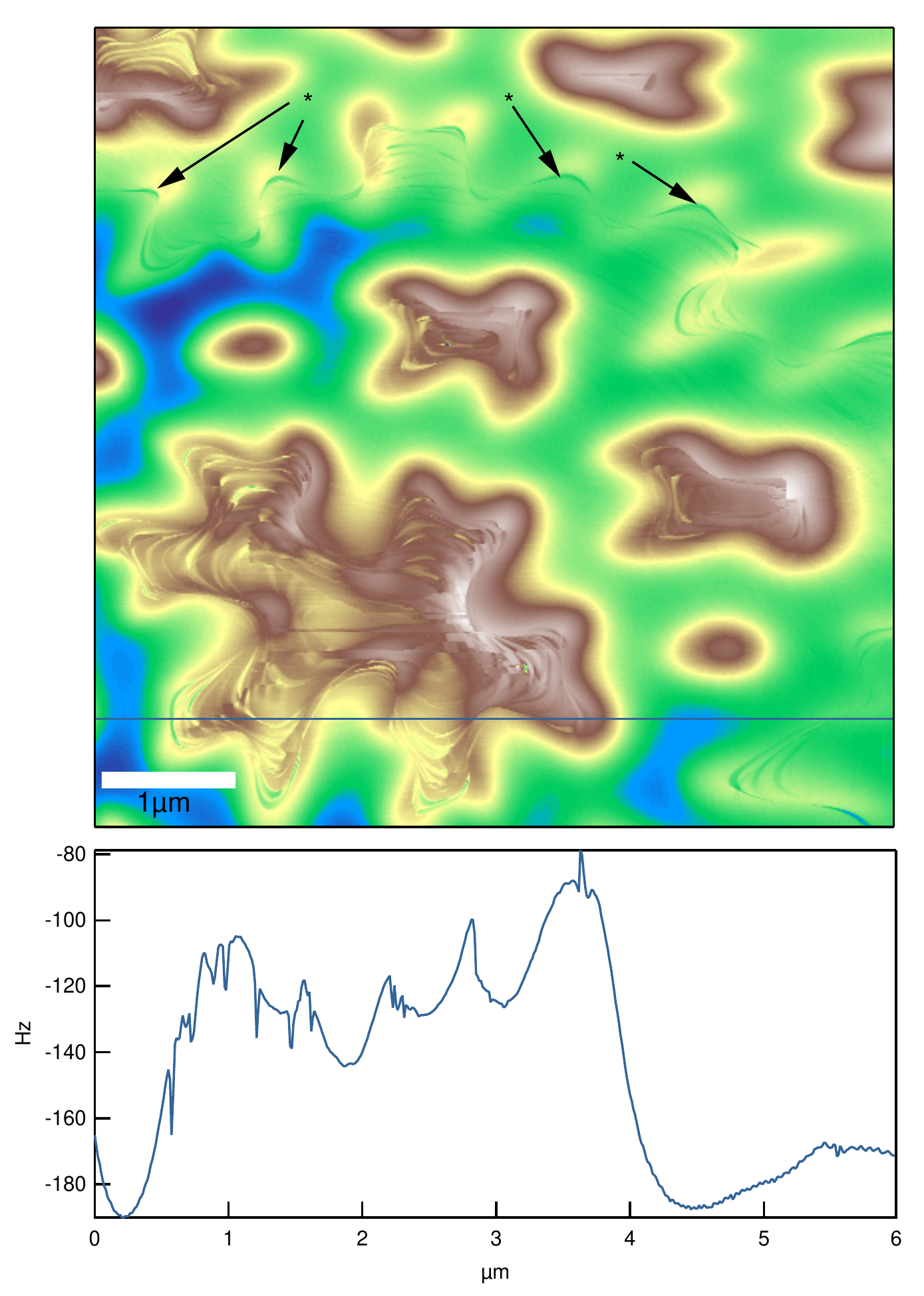}
\caption{(Color online) Top: A 6 $\mu$m * 6 $\mu$m magnetic frequency shift image of the surface of Nd$_2$Fe$_1$$_4$B.
Bottom: The z-profile of the blue line in the upper graph.
Scan Height: 300 nm.}
\label{fig2}
\end{figure}

\begin{figure}
\includegraphics[width=3.2in]{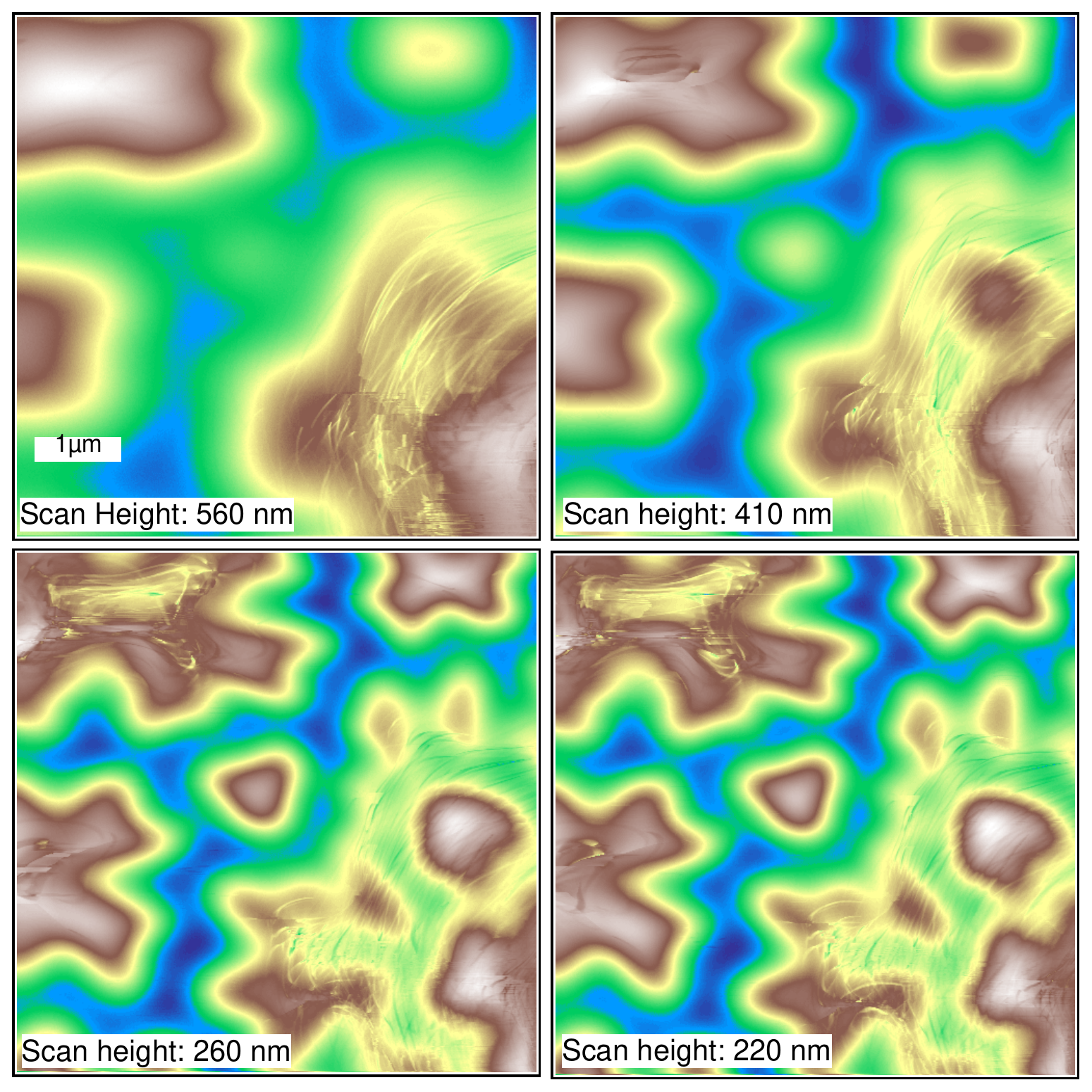}
\caption{(Color online) The same position on the sample is scanned at different tip-sample distances (6 $\mu$m * 6 $\mu$m. Scan Height are 560, 410, 260 and 220 nm respectively.).
}
\label{fig3}
\end{figure}

\begin{figure*}
\includegraphics[width=7in]{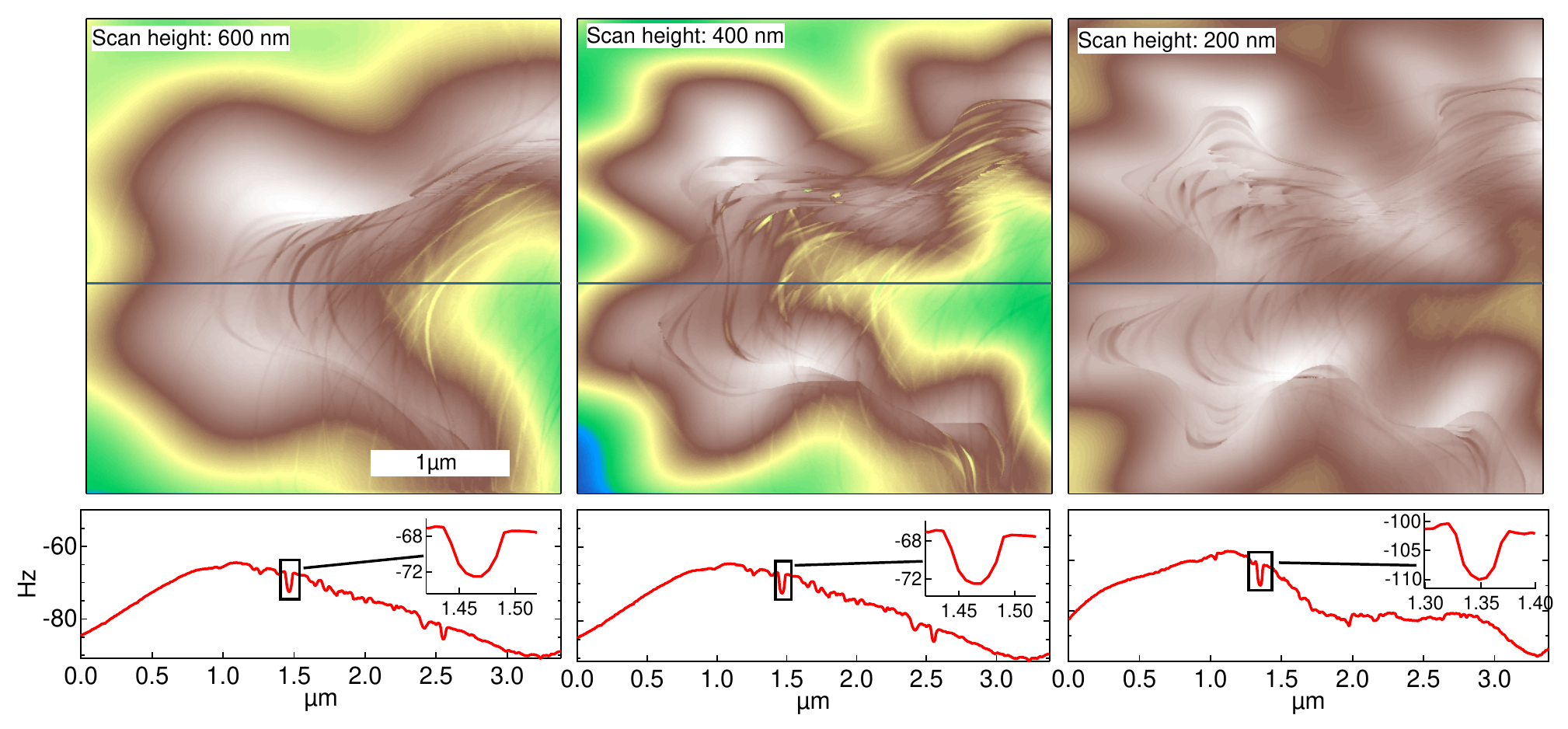}
\caption{(Color online) Top: The same position on the sample is scanned at a different tip-sample distance (6 $\mu$m * 6 $\mu$m). Bottom: The z-profile of the blue line in the upper graphs and a zoom-in of the boxed area in the profile. Scan Height: (From left to right) 600 nm, 400 nm, 200 nm}
\label{heightScan2}
\end{figure*}

Figure \ref{fig2} shows the magnetic domain structure of Nd$_2$Fe$_{14}$B measured using the magnetic AFM tip in non-contact mode at a tip-surface distance of 300 nm. Several interesting features are clearly visible. There are fairly weak, long and wavy domain walls that have been reported by previous MFM experiments\cite{Al-Khafaji98}, indicated by arrows in Fig. 2. The most pronounced features are star-like domains that are several $\mu$m across and these were previously observed via Kerr optical microscopy\cite{Kreyssig09}. With our enhanced resolution we can also see that the star shape objects are not single domains. Instead, they consist of a very complex network of much smaller, elongated magnetic nano domains seen as a pattern of thin brown lines in the yellow background of Fig. 2a and a very sharp series of dips in the profile shown in Fig. 2b.

In Fig. 3 we demonstrate how the imaging of the magnetic domains depends on the sample-tip distance. At high separation (e.g. $\sim$500 nm), the magnetic field from a large number of domains averages out, producing a smooth pattern of star-shaped objects that are a few $\mu$m across and similar to Kerr optical imaging. When the sample-tip distance is reduced, the magnetic field averaging effects are weaker and the tip begins to react to the presence of nano-size domains. This is best illustrated by following the evolution of the large domain in the upper left corner. 560 nm above the surface, this looks like a nice smooth single domain with round edges. At 410 nm, the tip begins to detect a variation of the magnetic field at the center of this object. At even smaller tip-surface separations (e. g. 220 nm), it is clear that this is not a single domain, instead it consists of fine network of nano-scale domains. This is shown in more detail in Fig. 4, where we focus on smaller area of the sample and part of a single micro domain. We can see that the overall shape of the micro domain is roughly similar, but a smaller surface-tip distance reveals a larger number of nano-domains. While certain, large features are visible for all three sample-tip separation, such as the wavy, yellow-brown edges of the star-shaped domains, others only appear at smaller scan heights. We can confirm that all micro domains look smooth and uniform at large scan heights. The smooth appearance of the star-shaped domains at large scan heights is simply a result of an averaging of the magnetic field away from the sample surface. At smaller scan heights, more and more nano-domains are revealed. Another, expected feature is observed by comparing the first two and last scans in Fig. 4. At large tip-sample distances, all the features are reproducible. Closer distances reveal finer detail, but the existing features are not modified. This is in contrast with small scan heights, where at 200nm, we observe that some features are significantly modified, while others remain unchanged. This is most likely a result of the magnetic field from the tip affecting the domain in the sample. This unwelcome phenomenon imposes a limit on the details that can be revealed by this technique.

\begin{figure*}
\includegraphics[width=7in]{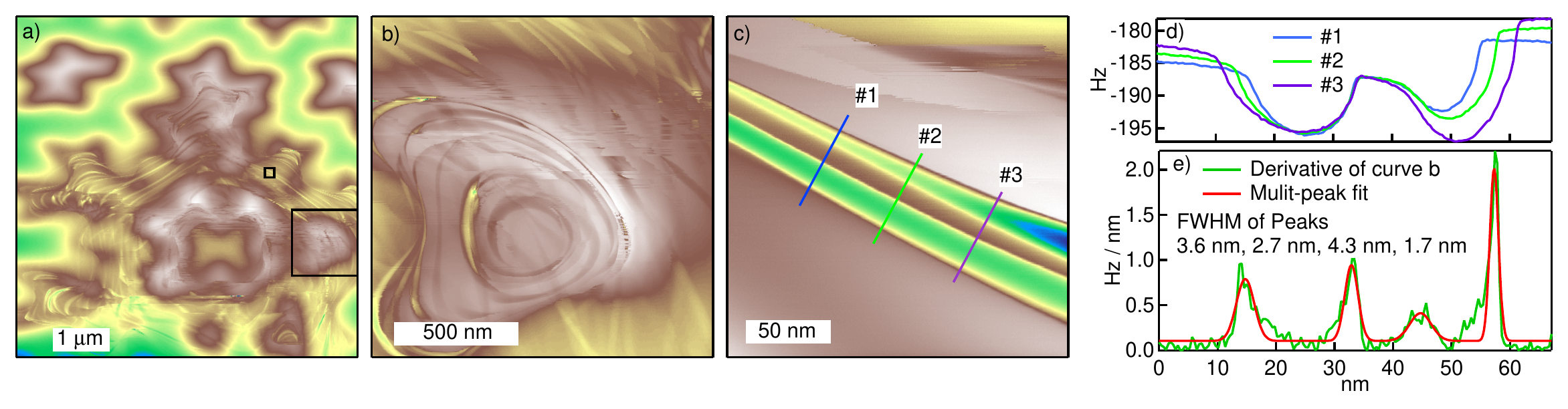}
\caption{(Color online) (a) A frequency shift image (6 $\mu$m * 5.5 $\mu$m) measured using magnetic tip 300nm above the surface of Nd$_2$Fe$_{14}$B. (b) A zoom-in from the larger box area marked in (a) (scanned area 1.37 $\mu$m * 1.37 $\mu$m). (c) A zoom-in from the smaller box area marked in (a) (scanned area 200 nm * 200 nm). (d) The z-profiles along the three cuts indicated in (c). (e) The  derivative of the curve \#2 in graph (d) and multi-peak fit.}
\label{fig6}
\end{figure*}

We will now examine the properties of the nano-domains in detail. In Fig. 5a we show the wide area scan of several star-shaped domains. We then focus on a smaller area that contains just a single object from which we select a very small 200x200 $\mu$m area shown in Fig. 5d. This shows three domains separated by areas of lower value of magnetic gradient seen as green/blue. Those features are very sharp even on tens of nm scale. We extract three cuts and examine the spatial variation of the cantilever frequency as a function of position along direction perpendicular to the direction of the domain walls. Those profiles are shown in Fig. 5e. The domain in the center is very narrow with width $\sim$10 nm. The low gradient areas separating the domains are slightly wider - about 20 nm across. To obtain information about the limits on the thickness of the domain walls we calculate the derivative of the profiles from Fig. 5e and plot these in panel f. While the peaks do not have an exact gaussian shape, an approximate fit yields widths of between 2-4~nm, which most likely reflects the spatial resolution of our instrument. 

\section{Conclusions}

We have studied the domain structure of Nd$_2$Fe$_{14}$B  using high resolution MFM. In addition to previously observed long, wavy nano-domains\cite{Al-Khafaji98}, we find that a star structure present at room temperature is formed from a complicated network of elongated (although much shorter) domains with typical widths of $\sim$20nm and a resolution-limited domain wall that is thiner than 2nm. We also found that most domains imaged at modest sample-tip distances are insensitive to the perturbation created by the magnetized tip. At smaller distances, however, a number of these domains change their appearance, which sets a limit on the experimental ability to measure their properties. Despite this, we show an excellent instrumental resolution  (better than 2 nm) and an imaging of magnetic features that can be achieved even at moderate scan heights.

\section{Acknowledgements}
This research was supported by the US Department of Energy, Office of Basic Energy Sciences, Division of Materials Sciences and Engineering (data acquisition and analysis). P. C. C., T. N. L and V. T. acknowledge support from the Critical Materials Institute, an Energy Innovation Hub funded by the U.S. Department of Energy, Office of Energy Efficiency and Renewable Energy, Advanced Manufacturing Office (sample growth). Ames Laboratory is operated for the US Department of Energy by the Iowa State University under Contract No. DE-AC02-07CH11358.


\begin{thebibliography}{99}


\bibitem{Herbst91} J. F. Herbst, 
Rev. Mod. Phys. {\bf 63}, 819 (1991).

\bibitem{Buschow86} K. H. J. Buschow, 
Mater. Sci. Rep. {\bf 1}, 1, (1986).

\bibitem{Lemke97} H. Lemke, T. Goddenhenrich, C. Heiden, and G. Thomas, 
IEEE Trans. Magn. {\bf 33}, 3685 (1997).

\bibitem{Grutter88} P. Grutter, E. Meyer, H. Heinzelmann, L. Rosenthaler, H.-R. Hidber, and H.-J. Guntherodt, J, 
Vac. Sci. Technol. A {\bf 6}, 279 (1988).

\bibitem{Neu04} V. Neu, S. Melcher, U. Hannemann, S. Fahler, and L. Schultz, 
Phys. Rev. B {\bf 70}, 144418  (2004).

\bibitem{Szmaja06} W. Szmaja, 
J. Magn. Magn. Mater. {\bf 301}, 546 (2006).

\bibitem{Lewis1}
L. H. Lewis, J.-Y. Wang, and P. Canfield, J. Appl. Phys. {\bf 83}, 6843 (1998); doi: 10.1063/1.367664

\bibitem{Lewis2} J.-Y. Wang, L. H. Lewis, D. O. Welch, and P. Canfield,
Mater. Charact. {\bf 41}, 201 (1998).

\bibitem{Kreyssig09} A. Kreyssig, R. Prozorov, C.D. Dewhurst, P.C. Canfield, R.W. McCallum, and A.I. Goldman,
Phys. Rev. Lett. {\bf 102}, 047204 (2009).

\bibitem{Al-Khafaji98} M. Al-Khafaji, W. M. Rainforth, M. R. J. Gibbs, J. E. L. Bishop, and H. A. Davies, 
J. Appl. Phys. {\bf 83}, 6411 (1998).


\bibitem{Pastushenkov97} Y.G. Pastushenkov, A. Forkl, and H. Kronmuller, 
J. Magn. Magn. Mater. {\bf 174}, 278 (1997).


\bibitem{crucible}
Paul C. Canfield and Ian R. Fisher, Journal of Crystal Growth {\bf 225}, 155Ð161 (2001)



\end{thebibliography}
\end{document}